\pdfoutput=1
\documentclass[aps,prl,reprint,nobibnotes]{revtex4-1}
\usepackage{graphicx}
\usepackage{amsmath}
\usepackage{amssymb}
\usepackage{amsfonts}
\usepackage{dcolumn}
\usepackage{dsfont}
\usepackage{latexsym}
\usepackage{rotating}
\usepackage{color}
\usepackage{latexsym}
\usepackage{bbm}
\usepackage{subfigure}
\usepackage{float}
\usepackage{epsfig}
\usepackage{psfrag}
\usepackage{natbib, hyperref}
\usepackage{bm}
\usepackage{amsthm}
\usepackage{eucal}
\usepackage{mathrsfs}
\usepackage{url}
\usepackage{braket}
\usepackage{ulem}

\usepackage[resetlabels]{multibib}  %adds supplementary citations
\newcites{Supp}{Supplementary~References}   %adds supplementary citations

\usepackage{color} % use colored text in latex

\usepackage{hyperref}
\hypersetup{
colorlinks=true,final=true,
        linkcolor=blue,
        citecolor=blue,
        filecolor=blue,
        urlcolor=blue,
}
\begin{document}

\title{Supercurrent as a Probe for Topological Superconductivity \\ in Magnetic Adatom Chains}
%\title{Spontaneous Spin Accumulation Around Adatom Chains \\ Hosting Majorana Bound States}

%\author{Narayan Mohanta}
%\email{narayan.mohanta@physik.uni-augsburg.de}
%\affiliation{Center for Electronic Correlations and Magnetism, Theoretical Physics III, Institute of Physics, University of Augsburg, 86135 Augsburg, Germany}
%\author{Arno P. Kampf}
%\affiliation{Center for Electronic Correlations and Magnetism, Theoretical Physics III, Institute of Physics, University of Augsburg, 86135 Augsburg, Germany}
%\author{Thilo Kopp}
%\affiliation{Center for Electronic Correlations and Magnetism, Experimental Physics VI, Institute of Physics, University of Augsburg, 86135 Augsburg, Germany}

\author{Narayan Mohanta$^{1,2}$, Arno P. Kampf$^{1}$, and Thilo Kopp$^{2}$}
\affiliation{Center for Electronic Correlations and Magnetism, $^{1}$Theoretical Physics III, $^{2}$Experimental Physics VI, Institute of Physics, University of Augsburg, 86135 Augsburg, Germany}

\begin{abstract}
A magnetic adatom chain, proximity coupled to a conventional superconductor with spin-orbit coupling, exhibits locally an odd-parity, spin-triplet pairing amplitude. We show that the singlet-triplet junction, thus formed, leads to a net spin accumulation in the near vicinity of the chain. The accumulated spins are polarized along the direction of the local $\mathbf{d}$-vector for triplet pairing and generate an enhanced  persistent current flowing around the chain. The spin polarization and the ``supercurrent'' reverse their directions beyond a critical exchange coupling strength at which the singlet superconducting order changes its sign on the chain. The current is strongly enhanced in the topological superconducting regime where Majorana bound states appear at the chain ends. The current and the spin profile offer alternative routes to characterize the topological superconducting state in adatom chains and islands.
\end{abstract}
           
\maketitle

Artificial lattices of magnetic adatoms, such as Fe, Cr or Gd, deposited on a spin-orbit coupled conventional superconductor, provide a versatile platform to realize Majorana bound states (MBS), locally accessible via scanning tunneling microscopy (STM)~\cite{PhysRevB.88.020407,Li2016,PhysRevLett.114.236803,Nadj-Perge602,Feldman2017,Pawlak2016,PhysRevLett.115.197204,Ruby_NanoLetters,2016arXiv160706353M}. Unlike the heterostructure-based schemes~\cite{PhysRevLett.100.096407,PhysRevLett.105.077001,Mourik1003,0295-5075-108-6-60001}, these adatom lattices can exhibit topological superconductivity (TSC) even in the absence of spin-orbit coupling (SOC) when the adatom moments within a chain form a real-space helix~\cite{PhysRevB.90.085124,PhysRevB.91.064505,PhysRevB.94.144509}. The adatoms act as magnetic impurities for the host superconductor and give rise to Yu-Shiba-Rushinov (YSR) states which hybridize to form a band dispersing within the energy gap of the superconductor~\cite{YU_acta1965,shiba_PTP1968,rushinov_JETP1969,1367-2630-17-2-023051}. In suitable parameter regimes, SOC induces chiral TSC in the YSR impurity band~\cite{PhysRevB.93.014517,PhysRevB.88.155420,PhysRevLett.114.106801,PhysRevB.90.235433,PhysRevB.93.094508,PhysRevB.94.100501,0953-8984-28-49-495703,PhysRevB.94.060501, PhysRevB.93.024507,PhysRevB.91.064502,PhysRevB.90.180503,PhysRevB.90.060507,PhysRevB.89.180505,PhysRevB.89.115109,Hui2015,PhysRevB.94.144509}. An adatom chain, therefore, mimics the physics of the one-dimensional Kitaev model~\cite{1063-7869-44-10S-S29} which contains isolated MBS at the ends of the chain. Experimental signatures for the possible existence of MBS on magnetic adatom chains were reported in recent STM experiments~\cite{Nadj-Perge602,Feldman2017,Pawlak2016,PhysRevLett.115.197204,Ruby_NanoLetters}. 

The broken inversion symmetry at an interface allows for a finite Rashba SOC. In the host $s$-wave superconductor, it combines with the local magnetic-exchange fields to produce spin-triplet, odd-parity, $\pm~p_x+~ip_y$-wave pairing at the adatom sites~\cite{2017arXiv170505378R}. Conversely, the exchange field at the adatom sites suppresses the $s$-wave order parameter. Therefore, the adatom sites possess an admixture of spin-singlet $s$-wave and spin-triplet $\pm~p_x+~ip_y$-wave pairing. The spin-triplet correlations decay rapidly away from the adatom sites.  An effective Josephson junction is, thus, formed in the vicinity of the adatom chain. A junction between a singlet and a triplet superconductor is predicted to accumulate spin because of the lifted spin-degeneracy of Andreev bound states (ABS)~\cite{PhysRevLett.101.187003}. Concomitantly, Rashba SOC generates a persistent current, carried by the YSR states, around the magnetic impurities; the currents flow orthogonal to the local spin polarization~\cite{PhysRevLett.115.116602}.

Here, we show that the two different mechanisms, \textit{viz.} the singlet-triplet Josephson junction and Rashba SOC, jointly cooperate to reinforce spin accumulation and spontaneous current flow around the adatom chains. Due to Rashba SOC, the accumulated spin provides an additional source for a circulating current around the adatom chain. The accumulated spins are polarized perpendicular to the direction of current flow and parallel to the local $\mathbf{d}$-vector of the triplet pairing amplitude. The $s$-wave pairing gap $ $ at the adatom site acquires a sign change as the YSR states undergo a parity-changing phase transition when the exchange coupling is tuned beyond a critical strength. The directions of both the spin-polarization and the current are reversed at the same exchange-coupling strength at which $\Delta_s$ changes sign. The current is significantly enhanced in the topological superconducting regime and, therefore, can be used as an alternative probe for the identification of TSC and the experimental verification of MBS.

We describe the adatom chain coupled to the host superconductor by the Hamiltonian $H_{\text{host}}+H_{\text{imp}}$. For a square lattice, $H_{\text{host}}$ is written as
\begin{align}
{H}_{\text{host}}&=-t \sum_{\langle ij \rangle,\sigma}c_{i \sigma}^{\dagger}c_{j\sigma}-\mu \sum_{i,\sigma}c_{i \sigma}^{\dagger}c_{i\sigma} \nonumber \\
& -i\alpha \sum_{\langle ij \rangle,\sigma, \sigma^{\prime}} (\boldsymbol{\sigma}\times \mathbf{d}_{ij})_{\sigma \sigma^{\prime}}^{z} c_{i\sigma}^{\dagger}c_{j\sigma^{\prime}} \nonumber \\
& +\sum_{i} (\Delta_{s}^{i} c_{i\uparrow}^{\dagger} c_{i\downarrow}^{\dagger}+\Delta_{s}^{i*} c_{i\downarrow}^{\dagger} c_{i\uparrow}^{\dagger}),
\label{H_host}
\end{align} 
where $t$ is the nearest-neighbor hopping amplitude of electrons, $\mu$ the chemical potential, and $\alpha$ the strength of Rashba SOC; $\mathbf{d}_{ij}$ denotes the unit vector between sites $i$ and $j$. $\Delta_s^{i}=-U_s \langle c_{i\uparrow} c_{i\downarrow} \rangle$ is the local $s$-wave pairing gap, with the strength $U_s$ of the onsite attraction. The adatom Hamiltonian $H_{\text{imp}}$ is expressed as
\begin{align}
H_{\text{imp}}=-J \sum_{i \in \mathbf{I}, \sigma, \sigma^{\prime}} (\mathbf{S}_i \cdot \boldsymbol{\sigma}^{z})_{\sigma \sigma^{\prime}} c_{i\sigma}^{\dagger}c_{i\sigma^{\prime}}
\label{H_imp}
\end{align} 
where $J$ is the strength of the exchange coupling between the adatom spin $\mathbf{S}_i$ and the conduction-electron spin, and $\mathbf{I}$ denotes the sub-lattice corresponding to the adatom-chain sites. We assume that all the adatom spins are identical and ferromagnetically aligned along the $z$-direction which is perpendicular to the surface of the superconductor, \textit{i.e.} $\mathbf{S}_i=S\hat{z}$. 

With this choice of the Hamiltonian, the local pairing amplitudes and other site-resolved observables are determined using the self-consistent Bogoliubov-de Gennes (BdG) formalism [see Supplemental Materials]. To study the induced triplet-pairing amplitude on and near the chain, we add an equal-spin, nearest-neighbor, triplet term $H_{\text{triplet}}=\sum_{\langle ij \rangle, \sigma} (\Delta_{\sigma \sigma}^{t,ij} c_{i\sigma}^{\dagger} c_{j\sigma}^{\dagger}+\Delta_{\sigma \sigma}^{t,ij*} c_{j\sigma}^{\dagger} c_{i\sigma}^{\dagger})$ where $\Delta_{\sigma \sigma}^{t,ij}=-U_t \langle c_{i\sigma} c_{j\sigma} \rangle$. $U_t$ is the strength of the attractive interaction in the triplet channel. Such an interaction is generated by parity fluctuations in the presence of Rashba SOC and ferromagnetic exchange fields~\cite{PhysRevLett.115.207002}. 

%Yet, as we show in the Supplemental Materials, our results are not sensitive to the specific choice of $U_t/U_s$, and for $U_t \ll U_s$ the triplet pairing amplitudes $\langle c_{i\sigma} c_{j\sigma} \rangle$ extrapolate to a finite $U_t$-independent value.

The total Hamiltonian $H_{\text{host}}+H_{\text{imp}}+H_{\text{triplet}}$ is diagonalized to obtain the pairing amplitudes $\Delta_s^i$ and $\Delta_{\sigma \sigma}^{t,ij}$ in the self-consistent BdG evaluation, performed on an $N \times N$ square lattice with an adatom chain of length $N_{\text{imp}}$ and open boundary conditions. Here, we show results for $U_t=U_s$, but we checked that the results, presented here, do not differ qualitatively for smaller values of $U_t$, even in the limit $U_t \rightarrow 0$. The even-parity triplet pairing amplitude necessarily vanishes because of the broken time-reversal symmetry (TRS) at the impurity sites~\cite{Ebisu_ptep2016}.

%---------------------------------------------
\begin{figure}[t]
\begin{center}
\epsfig{file=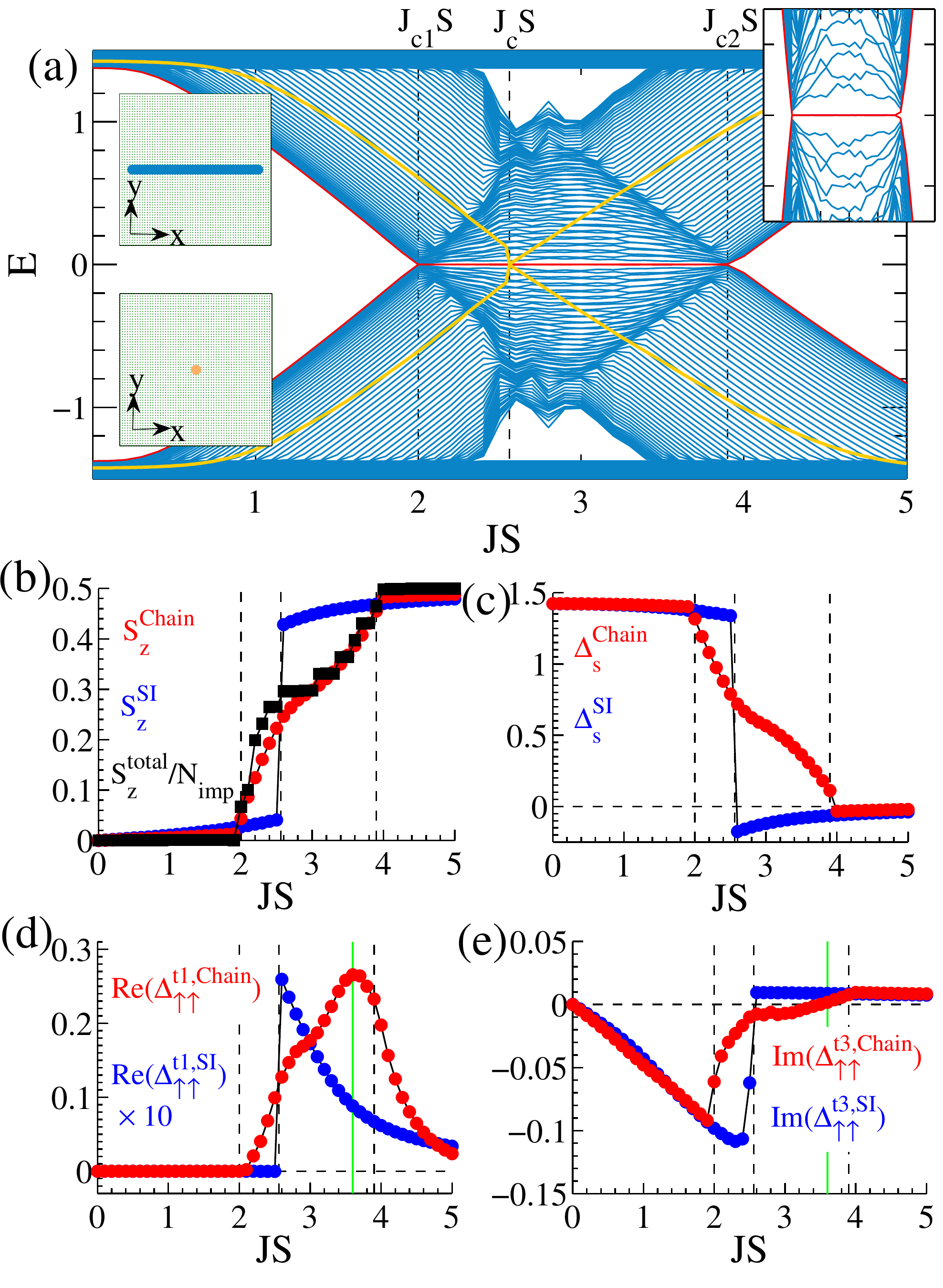,trim=0.0in 0.0in 0.0in 0.0in,clip=true, width=85mm}
 \vspace{-1em}
\caption{(Color online) (a) Variation of the BdG spectrum with impurity moment $JS$ for a chain of length $N_{\text{imp}}=50$, on a superconducting square lattice of size $61 \times 61$. The yellow lines are the pair of YSR states for a single magnetic impurity in the same superconducting host. The insets on the left show the adatom configurations. The pair of red lines at zero energy represents the MBS within the range $J_{c1}S~\leq~JS~\leq~J_{c2}S$. Right inset: expanded view of the spectrum for the chain within the topological range. (b), (c), (d) and (e) show, respectively, the $z$-component of spin-expectation value $S_z$, singlet pairing gap $\Delta_s$, and triplet pairing amplitudes $\Delta_{\uparrow \uparrow}^{t1}$, $\Delta_{\uparrow \uparrow}^{t3}$, averaged along the chain, (see main text) at the adatom sites or bonds with varying $JS$ for a chain and a single impurity. The black square symbols in (b) show the total $S_z$, divided by $N_{\text{imp}}=15$, for a $21\times21$ lattice. The vertical green lines in (d) and (e) denote the value of $JS$ at which Re($\Delta_{\uparrow \uparrow}^{t1, \text{Chain}}$) is maximum and Im($\Delta_{\uparrow \uparrow}^{t3, \text{Chain}}$) changes sign. Parameters taken: $t=1$, $\mu=0$, $\alpha=0.1$, and $U_s=3=U_t$.}
\label{Jvar}
\end{center}
%\vspace{-2em}
\end{figure}
%---------------------------------------------  
The pair of YSR states, originating from a single magnetic impurity in an $s$-wave superconductor, emerge from the continuum of bulk states and move symmetrically closer in energy within the bulk superconducting gap, as the impurity moment $JS$ is increased. A quantum phase transition occurs with respect to the total spin of the ground state when they cross zero energy at a critical value $J_cS$ ($=1/\pi N_0$, where $N_0$ is the normal-state density of states at the Fermi level~\cite{RevModPhys.78.373}). Beyond $J_cS$, the YSR states move apart and approach the bulk continuum states. For an adatom chain, the YSR states form a band and, there is a range $J_{c1}S \leq JS \leq J_{c2}S$ within which the YSR states remain close to zero energy. With a finite Rashba SOC and the inclusion of triplet pairing in the self-consistent calculation, there is a discontinuous change in energy of the YSR states for a single impurity at $J_cS=2.56$) rather than a smooth crossover, and for an adatom chain, the `zero-energy' MBS appear within the same range $J_{c1}S \leq JS \leq J_{c2}S$ (where, $J_{c1}S=2$ and $J_{c2}S=3.9$), as shown in Fig.~\ref{Jvar}(a) [see Supplemental Materials for details on MBS]. The critical numbers $J_{c}S$, $J_{c1}S$ and $J_{c2}S$ are modified in the presence of finite SOC~\cite{PhysRevLett.114.236804}. The pair of the zero-energy MBS is separated by a small energy gap, referred to as the minigap~\cite{PhysRevB.86.024504}, from the YSR band states as shown in the inset of Fig.~\ref{Jvar}(a). The spatial extent of the MBS, centered at the two ends, decays exponentially with distance, but still gives rise to a finite overlap at the center of the chain, creating a tiny hybridization gap ($\sim 10^{-4}$ for $N_{\text{imp}}=50$) which reduces with increasing chain length. 

For a single magnetic impurity, the zero-energy level-crossing of the YSR states leads to a quantum phase transition as a localized quasiparticle excitation, with spin opposite to the impurity spin, is spontaneously created, forming an antiferromagnetic bound state with the impurity spin~\cite{SakuraiPTP1970,PhysRevB.55.12648,RevModPhys.78.373,2017arXiv170505378R}. The transition at $J_{c}S$ changes the total $S_z$ of the ground state of the pairing Hamiltonian from $S_z=0$ to $\pm1/2$, depending on the sign of $J$ [see Supplemental Materials]. On the other hand, the $s$-wave pairing amplitude $\Delta_s$ at the impurity site sharply drops in magnitude at $J_cS$ and even encounters a sign change with respect to the bulk~\cite{RevModPhys.78.373,PhysRevB.95.104521}. 

For an adatom chain, the transition in the total $S_z$ of the conduction electrons $S_z=1/2\sum_{i,\sigma,\sigma'} \langle c_{i\sigma}^{\dagger} \sigma_{\sigma \sigma'}^z c_{i\sigma'} \rangle$ and the singlet pairing gap $\Delta_s$ take place within the same range $J_{c1}S \leq JS \leq J_{c2}S$ as depicted in Figs.~\ref{Jvar}(b) and \ref{Jvar}(c). With increasing $JS$, the total $S_z$ increases in discrete steps of height $1/2$~\cite{PhysRevB.73.224511}. The sign change in $\Delta_s$ occurs at $J_{c2}S$ when $S_z=1/2N_{\text{imp}}$. Also at $J_{c2}S$, the system becomes non-topological. 

We calculate the triplet pairing amplitudes $\Delta_{\sigma \sigma}^{t,ij}$ on the four bonds connected to the chain site $i$ \textit{viz.} $\Delta_{\sigma \sigma}^{t1}=\Delta_{\sigma \sigma}^{i,i+\hat{x}}$, $\Delta_{\sigma \sigma}^{t2}=\Delta_{\sigma \sigma}^{i,i-\hat{x}}$, $\Delta_{\sigma \sigma}^{t3}=\Delta_{\sigma \sigma}^{i,i+\hat{y}}$, and $\Delta_{\sigma \sigma}^{t4}=\Delta_{\sigma \sigma}^{i,i-\hat{y}}$. Due to the inherent $p$-wave symmetry, we obtain: $\Delta_{\sigma \sigma}^{t2}=-\Delta_{\sigma \sigma}^{t1}$ and $\Delta_{\sigma \sigma}^{t4}=-\Delta_{\sigma \sigma}^{t3}$. Furthermore, we find Im$(\Delta_{\sigma \sigma}^{t1})= $Re$(\Delta_{\sigma \sigma}^{t3})= 0$, Re$(\Delta_{\uparrow \uparrow}^{t1})=-\text{Re}(\Delta_{\downarrow \downarrow}^{t1})$ and Im$(\Delta_{\uparrow \uparrow}^{t3})=\text{Im}(\Delta_{\downarrow \downarrow}^{t3})$. This implies $-p_x+ip_y$-wave pairing for the $\uparrow \uparrow$ channel and $p_x+ip_y$-wave pairing for the $\downarrow \downarrow$ channel. The triplet pairing amplitudes reflect the broken TRS at the adatom sites; they decay rapidly away from the chain sites [see Supplemental Materials]. Figs.~\ref{Jvar}(d) and \ref{Jvar}(e) show the variation of the triplet pairing amplitudes with $JS$. For a single impurity, these amplitudes reveal sharp jumps at $J_cS$, along with a sign change in Im$(\Delta_{\sigma \sigma}^{t3})$. For the adatom chain, the transition proceeds within the range $J_{c1}S \leq JS \leq J_{c2}S$, along with a maximum in Re$(\Delta_{\sigma \sigma}^{t1})$ and a sign change in Im$(\Delta_{\sigma \sigma}^{t3})$ at $J_{cs}S=3.6$ (denoted by a green vertical line in Figs.~\ref{Jvar}(d)-(e)). Hence, at $J_{cs}S$, there is a change in pairing symmetry from $\pm p_x+ip_y$-wave to $\pm p_x-ip_y$-wave ($\pm$ refers to pairing in $\uparrow \uparrow$ and $\downarrow \downarrow$ channels, respectively).
%---------------------------------------------
\begin{figure}[t]
\begin{center}
\epsfig{file=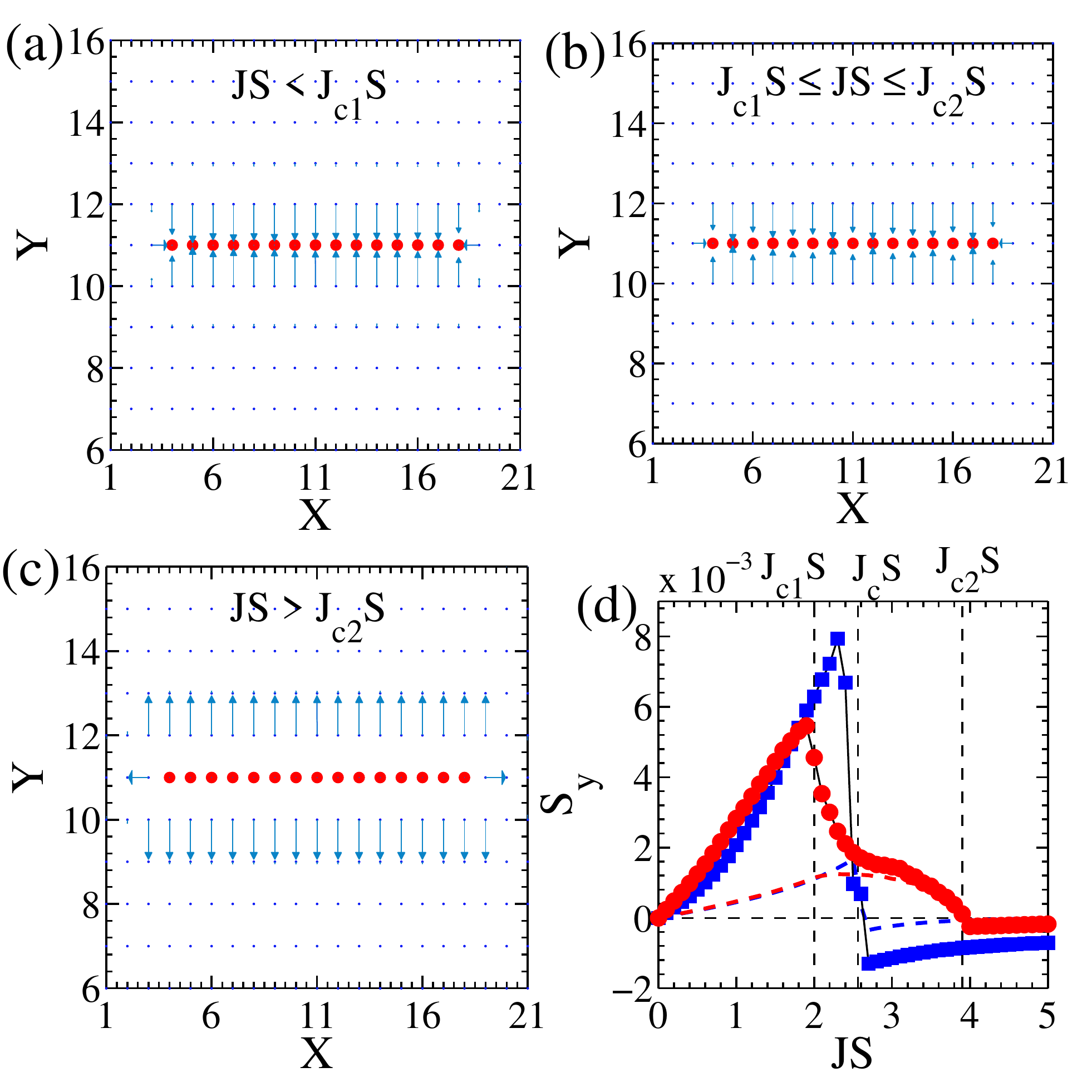,trim=0.0in 0.0in 0.0in 0.0in,clip=true, width=87mm}
\vspace{-2em}
\caption{(Color online) Profile of the accumulated spins near the adatom chain for (a) $JS=1 (<J_{c1}S)$, (b) $JS=3 (J_{c1}S \leq JS \leq J_{c2}S)$, and (c) $JS=5 (>J_{c2}S)$, calculated on a $21 \times 21$ lattice with $N_{\text{imp}}=15$. The red symbols mark the adatom sites. The length of the arrows is amplified for visualization purpose. (d) Variation of the spin-expectation value $S_y$, averaged over the sites just below the chain, with impurity moment $JS$ for finite $U_t$ and $U_t=0$ (solid symbols and dashed lines, respectively) in case of a single magnetic impurity (blue) and a chain (red). Parameters are the same as in Fig.~\ref{Jvar}.} 
\label{texture}
\end{center}
%\vspace{-2em}
\end{figure}
%---------------------------------------------

The triplet pairing is confined to the near vicinity of the adatom chain, and, therefore, an effective Josephson junction is formed between the chain region with pairing symmetry $s \pm p_x + ip_y$ and the surrounding host superconductor with $s$-wave symmetry. Enhancing the triplet-pairing amplitude at the junction also increases the magnitude of the accumulated spins which are polarized in the direction of the local $\mathbf{d}$-vector (in Balian-Werthamer representation) of the triplet pairing amplitude~\cite{PhysRevLett.101.187003,PhysRevB.80.024504} [see Supplemental Materials]. The spins near the chain point towards and away from the chain, respectively, for $JS$ below and above $J_{c2}S$. The reorientation of the $\mathbf{d}$-vector and the sign change in $\Delta_s$ both occur at the same critical impurity moment $J_cS$ for a single magnetic impurity. But for an adatom chain, the $\mathbf{d}$-vector flips its direction at $J_{cs}S$ while $\Delta_s$ changes sign at $J_{c2}S$. The resultant spin profile near the chain is plotted in Figs.~\ref{texture}(a)-(c), in the three regimes, \textit{i.e.} $JS<J_{c1}S$, $J_{c1}S \leq JS \leq J_{c2}S$, and $JS>J_{c2}S$. To get quantitative information about the spin polarization and the value of $JS$ at which the spins flip their directions, we calculate $S_y=1/2\langle \sigma_y \rangle$ at the neighboring sites below the adatom chain, with $U_t=0$ and with finite $U_t$, and plot the average value with increasing $JS$ in Fig.~\ref{texture}(d). We observe a slight variation of $S_y$ in the topological regime, that however decreases with increasing chain length. More importantly, $S_y$, with finite $U_t$, acquires significantly larger values below $J_{c1}S$ compared to that in the case of $U_t=0$. The amplitude of $S_y$ follows the behavior of Im$(\Delta_{\sigma \sigma}^{t3})$, suggesting a close connection between the accumulated spin and the triplet pairing amplitude. $S_y$ changes sign at $J_{c}S$ for a single impurity and at $J_{c2}S$ for an adatom chain. The sign change in Im$(\Delta_{\sigma \sigma}^{t3})$ at $J_{cs}S$ is not reflected in $S_y$ owing to the small amplitude of Im$(\Delta_{\sigma \sigma}^{t3})$ near $J_{cs}$. The exchange coupling jointly with Rashba SOC control the sign-change in $S_y$ at $J_{c2}S$ when $\Delta_s$ also changes sign. 

%---------------------------------------------
\begin{figure}[t]
\begin{center}
\epsfig{file=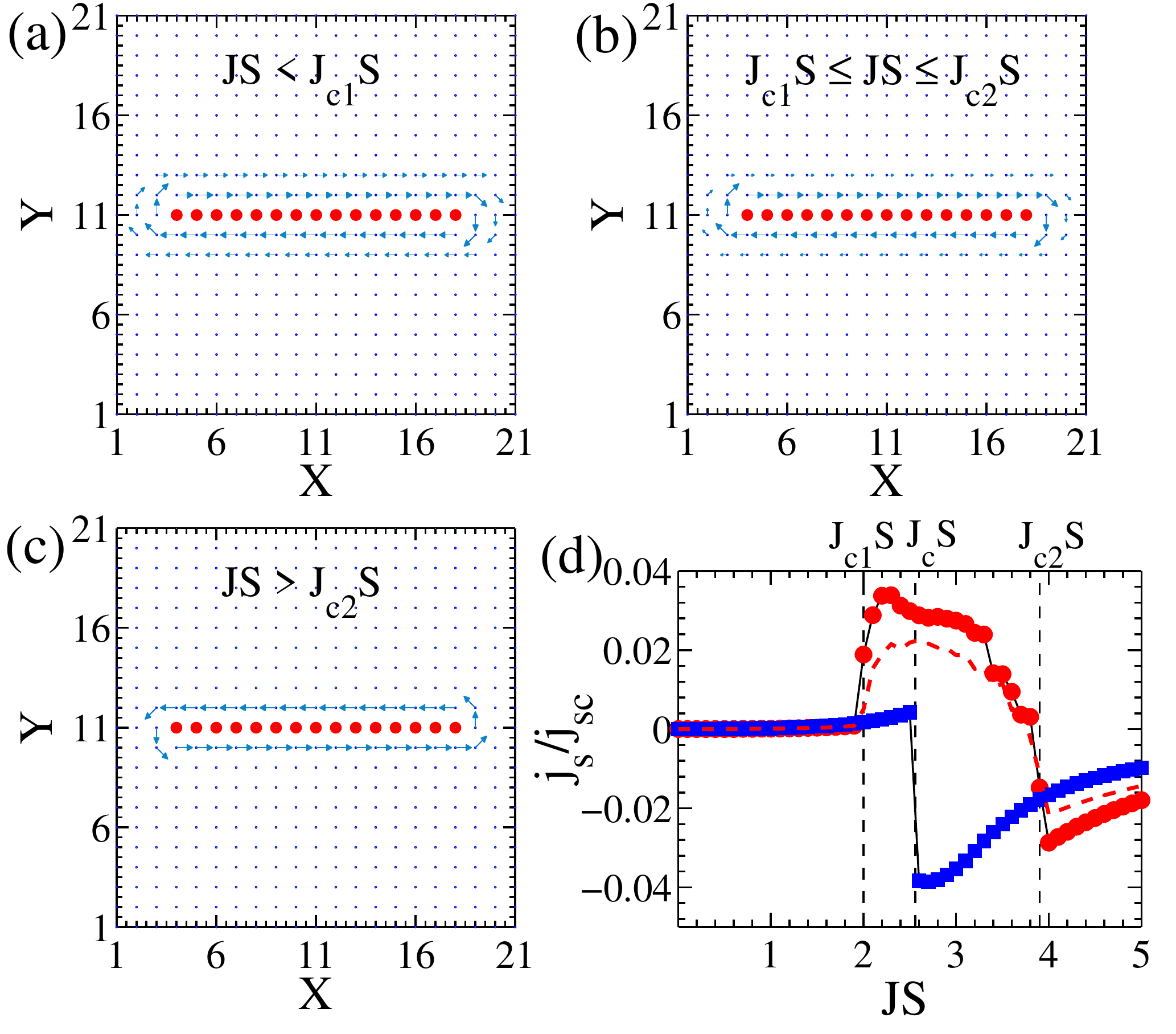,trim=0.0in 0.0in 0.0in 0.0in,clip=true, width=87mm}
\vspace{-2em}
\caption{(Color online) Current flow around the adatom chain for (a) $JS=1(<J_{c1}S)$, (b) $JS=3(J_{c1}S \leq JS \leq J_{c2}S)$, and (c) $JS=5(>J_{c2}S)$, calculated on a $21 \times 21$ lattice with $N_{\text{imp}}=15$. The red symbols mark the adatom sites. (d) Variation of the magnitude of the current $ \mathbf{j_{s}}$, normalized by the critical supercurrent for the host superconductor $\mathbf{j_{sc}}$, with the impurity moment $JS$ for a single magnetic impurity (blue squares) and an adatom chain (red circles). The dashed line shows the variation of the current with $JS$ for $U_t=0$. Parameters are the same as in Fig.~\ref{Jvar}.} 
\label{supercurrent}
\end{center}
%\vspace{-2em}
\end{figure}
%---------------------------------------------
Rashba SOC gives rise to an additional contribution to the paramagnetic current operator, the expectation value of which depends upon the magnetization profile. A magnetic impurity or a ferromagnetic island, proximity coupled to a superconductor, renders a finite spin polarization around the impurity or the island, giving rise to a SOC-driven current~\cite{PhysRevLett.115.116602,PhysRevB.92.214501}. Hence, the spin accumulation near the chain generates a current in the superconductor which circulates around the chain. Driving a current via magnetism is referred to as the magneto-electric effect~\cite{PhysRevB.72.172501}. The contribution to the current operator $\mathbf{j}_{s}^{SOC}=(j_{sx},~j_{sy})$ originating from Rashba SOC is given by
\begin{align*} 
j_{sx/sy}=\mp \alpha \frac{e}{\hbar}\sum_{\sigma,\sigma'} (\sigma_{y/x})_{\sigma \sigma'} (c_{i+\hat{x}/\hat{y},\sigma}^{\dagger}c_{i,\sigma'}-c_{i,\sigma'}^{\dagger}c_{i+\hat{x}/\hat{y},\sigma}).
\end{align*}
following the Peierls-factor derivation of Ref.~\cite{PhysRevB.47.7995}.
The rising of $S_{x/y}$ at sites next to the adatom chain, induced by the singlet-triplet junction, leads to an enhanced current around the chain as described in Fig.~\ref{supercurrent}. For a single magnetic impurity, the current $\mathbf{j}_{s}$ drops to a large negative value and changes its sense of circulation at $J_cS$ where $\Delta_s$ changes sign. However, for an adatom chain, $\mathbf{j}_{s}$ rises to higher values at $J_{c1}S$ (a few percent of the critical current $\mathbf{j}_{sc}=2e\Delta / \pi \hbar$ for the host superconductor~\cite{PhysRevB.49.6841}), decreases non-monotonically and finally reverses its propagation direction at $J_{c2}S$ as shown in Fig.~\ref{supercurrent}(a)-(d). The non-monotonic behavior of $\mathbf{j}_{s}$ between $J_{c1}S$ and $J_{c2}S$, including the steps, may be envisaged as a conjugate effect of (i) decrease in $S_y$ of the accumulated spins and (ii) the step-wise increase in total $S_z$ of the superconductor. The current is carried by the YSR states around the chain.

The magnetic-adatom chains have certainly received particular attention because of the existence of MBS at the chain ends. However, as our results demonstrate, the physics of the adatom chains contains far more intriguing aspects which originate from Rashba SOC. The accumulated spin polarization and the spin-triplet pairing amplitudes can possibly be detected by local spin-resolved probes with STM techniques~\cite{2017arXiv170100822G}, while the orbital magnetic moment of the persistent current around the adatom chains should be measurable using scanning SQUID sensors. The existence of the MBS, as inferred from STM experiments, appears still ambiguous primarily because of experimental challenges in disentangling the MBS from the YSR states near zero energy~\cite{1402-4896-2015-T164-014008}. The simultaneous detection of the circulating currents would therefore provide a further piece of evidence in support of TSC and the interpretation of the STM signals in favor of MBS.

For a particular choice of adatoms (Fe or Co) and their configuration on the surface, the parameters $J$ and $S$ are fixed and it is not possible to explore different phases in one experiment. The issue could be circumvented by applying a homogeneous magnetic field to rotate the spin polarization of the adatoms. In the current analysis, we assumed the adatom spins to point towards the $z$-direction \textit{i.e.} perpendicular to the surface of the superconductor. Since Rashba SOC breaks the SU(2) symmetry, the YSR states depend on the polar angle $\theta$ of the adatom spins and a topological phase transition can be induced by tuning $\theta$~\cite{PhysRevLett.114.236804,Loder2015}. A magnetic-field rotation experiment may thereby tune to the topological regime using the circulating current as a probe. The phenomena, identified here, are equally applicable to islands of adatoms where MBS convert to edge-bound modes.

The authors acknowledge discussions with Dirk Morr, Stephan Rachel, and Titus Neupert. This work was supported by the Deutsche Forschungsgemeinschaft through TRR~80. Computations were performed at the Leibniz-Rechenzentrum in Munich.

%additional citation~\cite{PhysRevB.92.014513}

\bibliographystyle{h-physrev}
%\bibliographystyle{plainnat}
%\bibliography{Ref_Shiba}

%%%%%%%%%% Merge with supplemental materials %%%%%%%%%%
\newpage
%\pagebreak
\newpage
\clearpage
\widetext
\begin{center}
%\textbf{\large Supplemental Materials: Spontaneous Spin Accumulation Around Adatom Chains Hosting Majorana Bound States}
\textbf{\large Supplemental Materials: Supercurrent as a Probe for Topological Superconductivity in Magnetic Adatom Chains}
\end{center}
%%%%%%%%%% Merge with supplemental materials %%%%%%%%%%
%%%%%%%%%% Prefix a "S" to all equations, figures, tables and reset the counter %%%%%%%%%%
\setcounter{equation}{0}
\setcounter{figure}{0}
\setcounter{table}{0}
\setcounter{page}{1}
\makeatletter
\renewcommand{\theequation}{E\arabic{equation}}
\renewcommand{\thefigure}{F\arabic{figure}}
\renewcommand{\bibnumfmt}[1]{[R#1]}
\renewcommand{\citenumfont}[1]{R#1}
%%%%%%%%%% Prefix a "S" to all equations, figures, tables and reset the counter %%%%%%%%%%

%======================EXAMPLE========================================================
%\section{Section 1}
%Copy and paste your Supplemental Materials text here \cite{S_RefA}, blah, blah, blah, blah, blah, blah, ...
%\begin{equation}
%  i\hbar\frac{\partial}{\partial t}\psi(x,t) = -\frac{\hbar^2}{2m}\frac{\partial^2}{\partial x^2}\psi(x,t) + V(x,t) \psi(x,t)
%\end{equation}

%\begin{thebibliography}{11}
%\bibitem{S_RefA} A. Someone, C. Someone, D. Someone, Phys. Rev. Lett. {\bf 11}, 1111 (1911).
%\end{thebibliography}
%======================================================================================

\section{S1. Self-consistent Bogoliubov-de Gennes (B{d}G) formalism}

The total Hamiltonian $H_{\text{tot}}=H_{\text{host}}+H_{\text{imp}}+H_{\text{triplet}}$ for a spin-orbit coupled superconductor with an adatom chain is written in a square lattice as
\begin{eqnarray}
H_{\text{tot}}&=& -t {\sum_{ \langle ij \rangle ,\sigma}}c_{i{\sigma}}^{\dagger} c_{j{\sigma}} - {\mu} \sum_{i,\sigma}c_{i{\sigma}}^{\dagger} c_{i{\sigma}}  - J\sum_{i\in \mathbf{I},\sigma,{\sigma}'}(\mathbf{S}_i \cdot \boldsymbol{\sigma}^{z})_{\sigma{\sigma}'}c_{i{\sigma}}^{\dagger}c_{i{\sigma}'} 
 -i\alpha \sum_{\langle ij \rangle, \sigma,{\sigma}'} {({\boldsymbol{\sigma}}{{\times}\mathbf{d}_{ij})}^{z}_{{\sigma}{\sigma}'}} c_{i{\sigma}}^{\dagger} c_{j{\sigma}'} \notag\\
&&+ \sum_{i} (\Delta_ic_{i\uparrow}^{\dagger}c_{i\downarrow}^{\dagger} + \Delta_i^{*}c_{i\downarrow}^{\dagger}c_{i\uparrow}^{\dagger}) 
+\sum_{\langle ij \rangle, \sigma} (\Delta_{ij}^{\sigma \sigma} c_{i\sigma}^{\dagger} c_{j\sigma}^{\dagger}+\Delta_{ij}^{\sigma \sigma *} c_{j\sigma}^{\dagger} c_{i\sigma}^{\dagger})
\label{H_supple}
\end{eqnarray}
\noindent where $t$ is the hopping amplitude, $\mu$ the chemical potential, $J$ the exchange coupling strength of the adatom spin $\mathbf{S}_i$ (assumed to be polarized along the $z$-direction) with the conduction electrons. $\mathbf{I}$ denotes the sub-lattice of the adatom chain, $\alpha$ is the strength of Rashba spin-orbit coupling, and $\mathbf{d}_{ij}$ the unit vector between nearest-neighbor sites $i$ and $j$. $\Delta_{i}=-U_s\langle c_{i\uparrow}c_{i\downarrow} \rangle$ is the onsite singlet pairing gap with the attractive pairing potential $U_s$ and $\Delta_{ij}^{\sigma \sigma}=-U_t \langle c_{i\sigma} c_{j\sigma} \rangle$ denotes the nearest-neighbor equal-spin pairing gap with the attractive triplet pairing potential $U_t$.

The above Hamiltonian is diagonalized using the Bogoliubov-Valatin transformation:
\begin{align}
c_{i\sigma}=\sum_{n,\sigma^{\prime}} \left[ u_{n\sigma \sigma^{\prime}}^{i} \gamma_{n\sigma^{\prime}} + v_{n\sigma \sigma^{\prime}}^{i*} \gamma_{n\sigma^{\prime}}^{\dagger} \right]
\label{bdg_trans}
\end{align}
where  $\gamma_{n\sigma^{\prime}}^{\dagger}$ ($\gamma_{n\sigma^{\prime}}$) is a fermionic operator which describes the creation (destruction) of a BdG state with spin $\sigma^{\prime}$ in the $n^{th}$ eigenstate of $H_{tot}$; $u_{n\sigma\sigma^{\prime}}^i$ and $v_{n\sigma\sigma^{\prime}}^i$ are the particle and hole amplitudes, respectively. After diagonalization, $H_{\text{tot}}$ is rewritten in terms of $u_{n\sigma\sigma^{\prime}}^i$ and $v_{n\sigma\sigma^{\prime}}^i$. It is convenient, therefore, to remove the index $\sigma^{\prime}$ attached to these variables, whereby the summation over $\sigma'$ is implicitly subjoined in the sum over n. The BdG equations can finally be written, in the matrix form, as
\begin{align}
\sum_i \begin{pmatrix} \begin{array}{cccc} \Gamma_{ij}^{\uparrow\uparrow} & \Gamma_{ij}^{\uparrow\downarrow} & \Delta_{ij}^{\uparrow \uparrow} & \Delta_{ij}^{\uparrow \downarrow} \\ \Gamma_{ij}^{\downarrow\uparrow} & \Gamma_{ij}^{\downarrow\downarrow} & -\Delta_{ij}^{\downarrow \uparrow} & \Delta_{ij}^{\downarrow \downarrow} \\ \Delta_{ij}^{\uparrow \uparrow *} & -\Delta_{ij}^{\downarrow \uparrow*} & -\Gamma_{ij}^{\uparrow\uparrow *} & -\Gamma_{ij}^{\uparrow\downarrow *}\\\Delta_{ij}^{\uparrow \downarrow*} & \Delta_{ij}^{\downarrow \downarrow *} & -\Gamma_{ij}^{\downarrow\uparrow *} & -\Gamma_{ij}^{\downarrow\downarrow *}\end{array} \end{pmatrix}
\begin{pmatrix} \begin{array}{c} u_{n\uparrow}^j \\ u_{n\downarrow}^j \\ v_{n\uparrow}^j \\ v_{n\downarrow}^j \end{array} \end{pmatrix}=E_n \begin{pmatrix} \begin{array}{c} u_{n\uparrow}^j \\ u_{n\downarrow}^j \\ v_{n\uparrow}^j \\ v_{n\downarrow}^j \end{array} \end{pmatrix}
\label{BdG_eq}
\end{align}
\noindent where $\Gamma_{ij}^{\sigma\sigma^{\prime}}=-t(1-\delta_{ij})\delta_{\sigma \sigma^{\prime}}-\mu \delta_{ij} \delta_{\sigma \sigma^{\prime}}-J(\mathbf{S}_i \cdot \boldsymbol{\sigma}^{z})_{\sigma\sigma^{\prime}}\delta_{ij}(1-\delta_{\sigma\sigma^{\prime}}) -i\alpha ({\boldsymbol{\sigma}}{\times}\mathbf{d}_{ij})_{\sigma \sigma^{'}}^{z}(1-\delta_{ij})(1- \delta_{\sigma \sigma^{\prime}})$ ($\delta$'s are the Kronecker's delta functions), and $E_n$ is the energy of $n^{th}$ eigenstate. For a square lattice of size $N \times N$, the above matrix is of dimension $4N^2 \times 4N^2$.

The local pairing amplitudes and the magnetization components are written in terms of the particle and hole amplitudes as 
\begin{align}
&\Delta_i=-U_s\langle c_{i\uparrow}c_{i\downarrow}\rangle=-U_s\sum_{n}  \Big[ u_{n\uparrow}^i v_{n\downarrow}^{i*}(1-f(E_n))+u_{n\downarrow}^i v_{n\uparrow}^{i*}f(E_n) \Big], \nonumber \\
&\Delta_{ij}^{\sigma \sigma}=-U_t \langle c_{i\sigma}c_{j\sigma}\rangle=-U_t\sum_{n} \Big[ u_{n\sigma}^i v_{n\sigma}^{j*}(1-f(E_n))+u_{n\sigma}^i v_{n\sigma}^{j*}f(E_n) \Big], \nonumber \\
&S_{xi}=\frac{1}{2}\langle c_{i\uparrow}^{\dagger}c_{i\downarrow}+c_{i\downarrow}^{\dagger}c_{i\uparrow} \rangle=\frac{1}{2}\sum_{n} \Big[ (u_{n\uparrow}^{i*}u_{n\downarrow}^i+u_{n\downarrow}^{i*}u_{n\uparrow}^i) f(E_n)+(v_{n\uparrow}^{i}v_{n\downarrow}^{i*}+v_{n\downarrow}^{i}v_{n\uparrow}^{i*})(1-f(E_n)) \Big], \nonumber \\
&S_{yi}=\frac{-i}{2}\langle c_{i\uparrow}^{\dagger}c_{i\downarrow}-c_{i\downarrow}^{\dagger}c_{i\uparrow} \rangle=\frac{1}{2}\sum_{n} \Big[ (u_{n\uparrow}^{i*}u_{n\downarrow}^i-u_{n\downarrow}^{i*}u_{n\uparrow}^i) f(E_n)+(v_{n\uparrow}^{i}v_{n\downarrow}^{i*}-v_{n\downarrow}^{i}v_{n\uparrow}^{i*})(1-f(E_n)) \Big], \nonumber \\
&S_{zi}=\frac{1}{2}\langle c_{i\uparrow}^{\dagger}c_{i\uparrow}-c_{i\downarrow}^{\dagger}c_{i\downarrow} \rangle=\frac{1}{2}\sum_{n} \Big[ (|u_{n\uparrow}^{i*}|^2 - |u_{n\downarrow}^{i*}|^2) f(E_n)+(|v_{n\uparrow}^{i*}|^2 - |v_{n\downarrow}^{i*}|^2) (1-f(E_n)) \Big],
\label{gap}
\end{align}
\noindent where $f(E_n)$ is the Fermi-Dirac distribution function, entering in the above equations via the relations: $\langle \gamma_n^{\dagger} \gamma_n \rangle=f(E_n)$ and $\langle \gamma_n \gamma_n^{\dagger} \rangle=1-f(E_n)$. In what follows, iterations are performed using Eq.~(\ref{BdG_eq}) and Eqs.~(\ref{gap}) until self-consistency is achieved at every sites, and finally the local order parameters are computed using Eqs.~(\ref{gap}).

The local density of states (LDOS) is given by
\begin{align}
\rho_i(E)=\frac{1}{N^2} \sum_{n,\sigma}  \Big[ |u_{n\sigma}^{i}|^2\delta(E-E_n)+|v_{n\sigma}^{i}|^2\delta(E+E_n)  \Big],
\end{align}
\noindent where $\delta(E \pm E_n)$ are the Dirac delta functions which are approximated by Gaussian functions in the numerical evaluations.

\section{S2. Majorana bound states in the topological superconducting phase}
\begin{figure}[h!]
\begin{center}
\epsfig{file=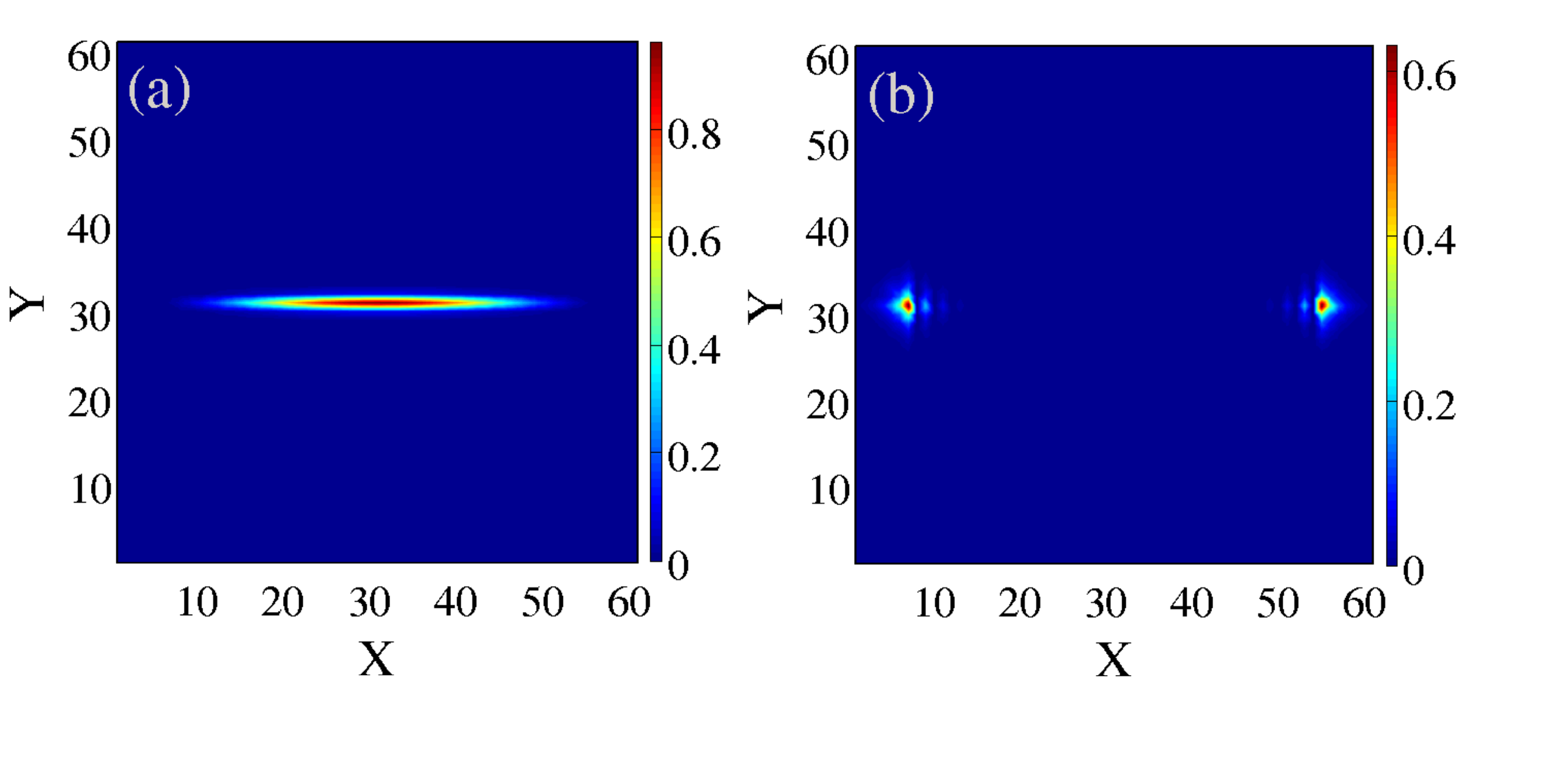,trim=0.0in 0.0in 0.0in 0.1in,clip=true, width=140mm}
\caption{The LDOS profile of an adatom chain of length $N_{\text{imp}}=50$, on a superconducting square lattice of size $61 \times 61$, corresponding to the pair of lowest energy eigenvalues for exchange-coupling strength (a) $JS=1$ (non-topological phase), and $JS=3$ (topological phase). The sharp peaks at the ends of the chain in (b) represent the zero-energy localized MBS. Parameters are the same as in Fig.~\ref{Jvar} in the main text.} 
\label{ldos}
\end{center}
\end{figure}

As made evident in Fig.~\ref{Jvar} in the main text, the magnetic adatom chain exhibits topological superconductivity within the range $J_{c1}S \leq JS \leq J_{c2}S$. To verify the Majorana bound states (MBS) appearing within this range of $JS$, we plot the LDOS profile corresponding to the lowest pair of energy eigenvalues for a square lattice of size $61 \times 61$ with a $50$-sites long adatom chain for $JS=1$ (non-topological phase) in Fig.~\ref{ldos}(a) and for $JS=3$ (topological phase) in Fig.~\ref{ldos}(b).
%---------------------------------------------
%---------------------------------------------
For $JS=1$, the LDOS profile is centered in the middle of the chain indicating that the Yu-Shiba-Rushinov (YSR) states are extended within the impurity chain. Instead, for $JS=3$, the lowest-energy pair ($\pm2.8\times10^{-4}$) comes close to zero and the LDOS is concentrated at the two ends of the adatom chain. These sharp features at the chain-ends are the signatures of localized MBS. The tiny energy gap ($\sim 10^{-4}$ for $N_{\text{imp}}=50$) between the two MBS arises because the two MBS which decay exponentially with distance, but hybridize in the middle of the chain. This hybridization gap reduces with increasing chain length.

%\newpage 
\section{S3. Spin-triplet pairing at the adatom chain}
\begin{figure}[htb!]
\begin{center}
\epsfig{file=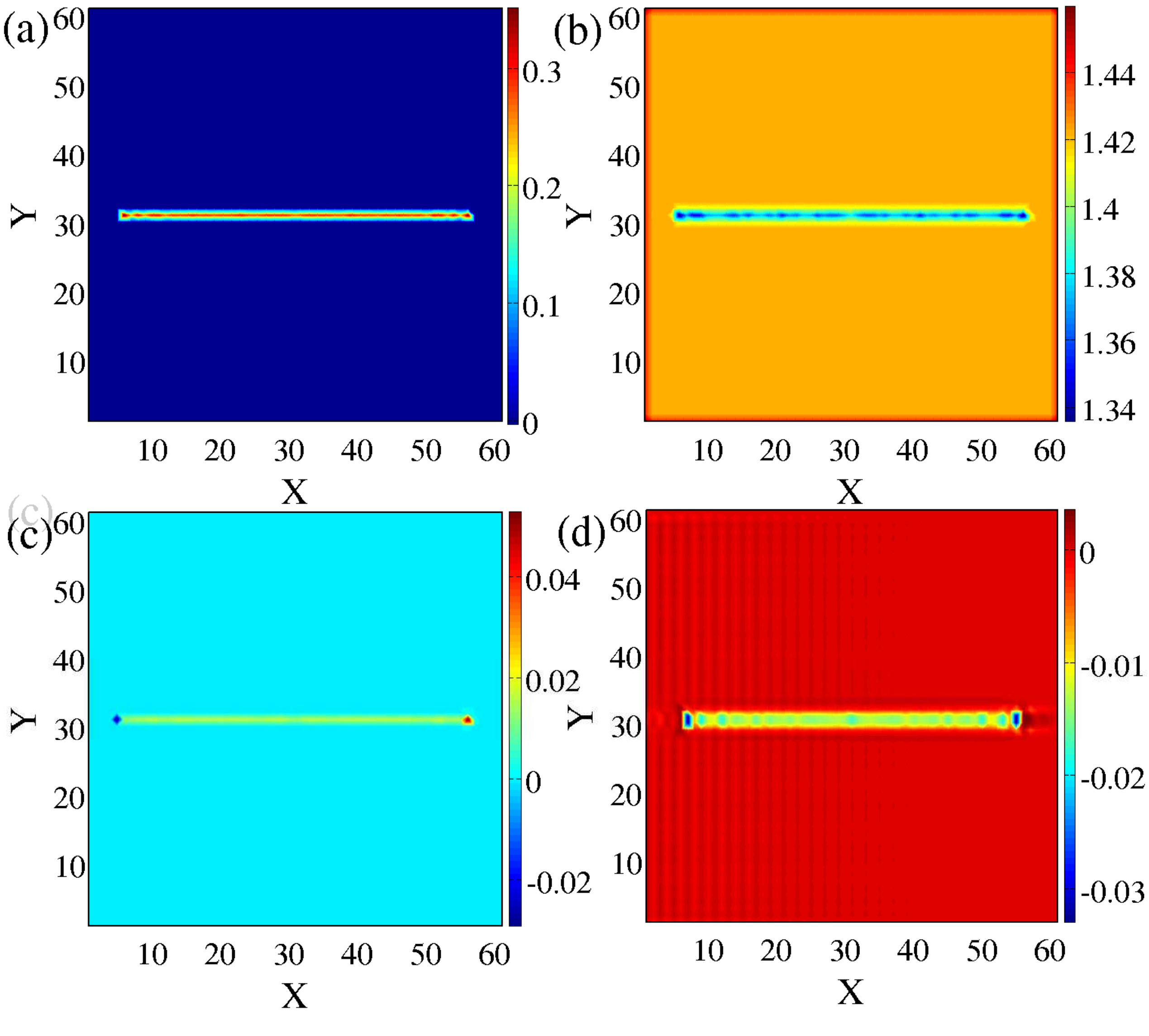,trim=0.0in 0.0in 0.0in 0.0in,clip=true, width=140mm}
\caption{The spatial profile of (a) $z$-component of spin-expectation value $S_{z}$, (b) singlet $s$-wave superconducting order parameter $\Delta_s$, (c) real part of triplet pairing gap for the bond along $+x$ direction, Re$(\Delta_{\uparrow \uparrow}^{t,1})$, and (d) imaginary part of triplet pairing gap for the bond along $+y$ direction, Im$(\Delta_{\uparrow \uparrow}^{t,3})$, calculated for $JS=3$ (topological superconducting phase). A square lattice of size $61 \times 61$ was used for the calculation with impurity-chain length $N_{\text{imp}}=50$ and parameters of Fig.~\ref{Jvar} in the main text.} 
\label{pairing_profile}
\end{center}
\end{figure}

To study the spatial confinement of the induced triplet pairing amplitude, we plot in Fig.~\ref{pairing_profile} the profile of the magnetization, the singlet and triplet pairing amplitudes in the two-dimensional lattice plane. 
%---------------------------------------------
%---------------------------------------------
We find that the induced triplet pairing is confined to the vicinity of the adatom chain and die off quickly at sites further away from the chain. The $s$-wave pairing gap $\Delta_s$ is significantly suppressed because of the local magnetization along the chain, as shown in Fig.~\ref{pairing_profile}(a)-(b). Interestingly, the triplet order parameters Re$(\Delta_{\uparrow \uparrow}^{t,1})$ and Im$(\Delta_{\uparrow \uparrow}^{t,3})$ reveal maximum amplitudes at the ends of the chain only in the topological superconducting regime, as shown in Fig.~\ref{pairing_profile}(c)-(d). It was, in fact, proposed that MBS on the boundary of topological superconductors have universal spin-triplet correlations~\cite{Liu_PRB2015}. There is a sign change in Re$(\Delta_{\uparrow \uparrow}^{t,1})$ at the two ends of the chain indicating the realization of chiral $p$-wave triplet pairing. The results demonstrate that a mixed singlet-triplet pairing is realized in the chain and, thus an effective Josephson junction is formed in the vicinity of the chain. The consequences of the  Josephson junction is discussed in the main text.

\begin{figure}[htb!]
\begin{center}
\epsfig{file=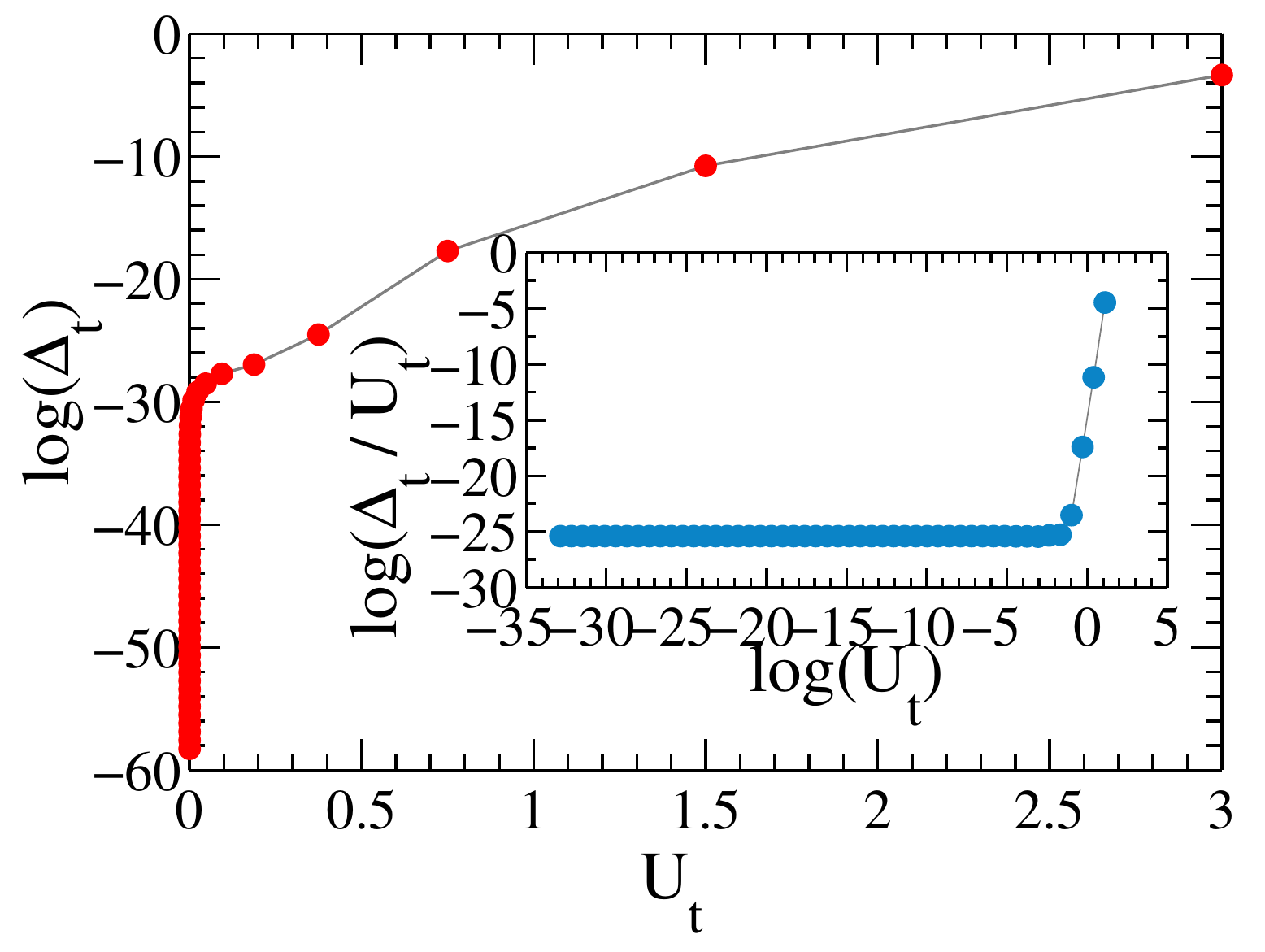,trim=0.0in 0.0in 0.0in 0.0in,clip=true, width=100mm}
\caption{Variation of the triplet order parameter $\Delta_t$ in natural logarithmic scale with the attractive interaction strength $U_t$ for impurity moment $JS=3$. Inset show the variation of log$(\Delta_t/U_t)$ with log$(U_t)$. Other parameters are the same as in Fig.~\ref{pairing_profile}.} 
\label{triplet_log_plot}
\end{center}
\end{figure}

For a conventional superconductor, the superconducting critical temperature $T_c$ and the pairing amplitude $\Delta$ are related to the attractive interaction strength $U$ via $k_{B}T_c =1.13 E_D e^{-1/N(0)U}$ and $\Delta (T=0)=1.76k_{B}T_c$, where $E_D$ is a cutoff energy for pairing and $N(0)$ is the density of states at the Fermi level. In order to check the nature of the induced triplet pairing in the adatom chain, we plot the logarithm of the triplet pairing amplitude $\Delta_t=\text{Re}(\Delta_{\uparrow \uparrow}^{t1})$ with $U_t$ in Fig.~\ref{triplet_log_plot}. 
%---------------------------------------------
%---------------------------------------------
In the $U_t \rightarrow 0$ limit, log($\Delta_t$) approaches a finite, small number. The results establish that the induced triplet pairing on the adatom chain is of conventional character.  

\begin{figure}[htb!]
\begin{center}
\epsfig{file=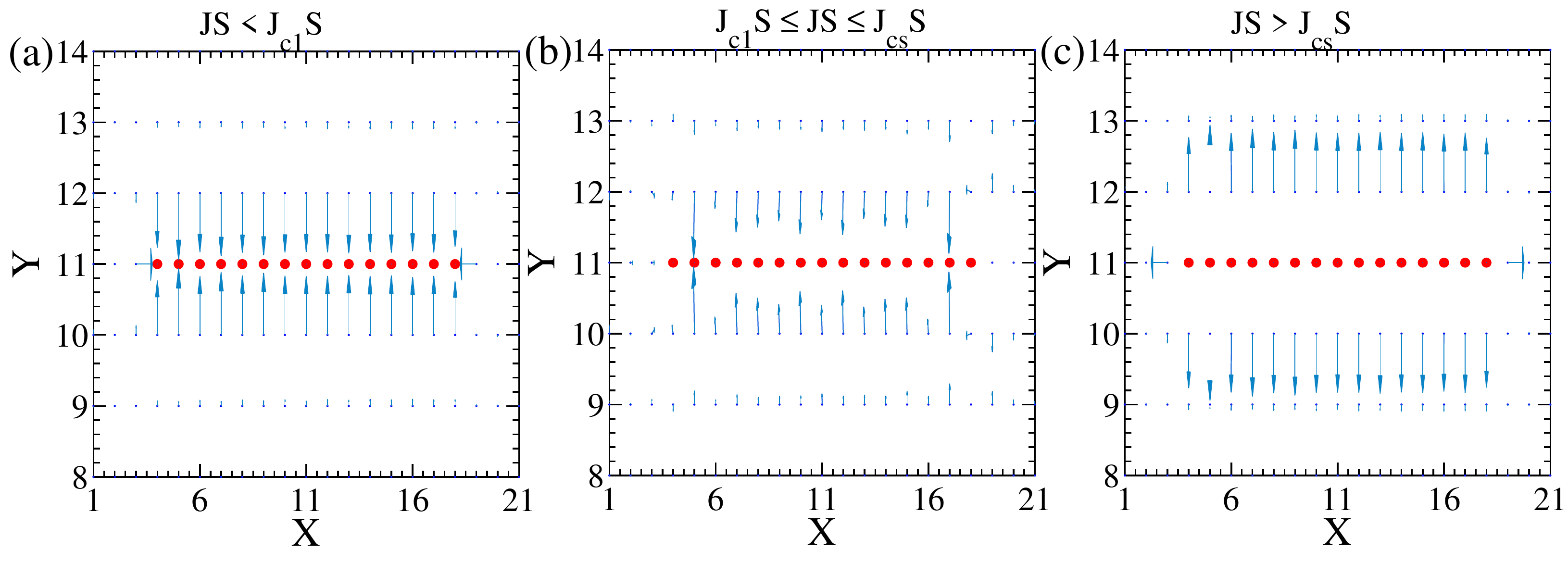,trim=0.0in 0.0in 0.0in 0.0in,clip=true, width=140mm}
\caption{The real-space profile of $\mathbf{d}$-vector of the triplet pairing amplitude for (a) $JS=1$ (trivial phase), (b) $JS=3$ (topological superconducting phase) and (c) $JS=5$ (trivial phase). Other parameters are the same as in Fig.~\ref{pairing_profile}.} 
\label{dvector}
\end{center}
\end{figure}

To identify the connection between the polarization of the accumulated spins near the chain and the induced triplet pairing, we plot the local $\mathbf{d}$-vector of the triplet pairing amplitude in real space in Fig.~\ref{dvector}. In Balian-Werthamer representation, the pairing matrix can be written as
\begin{align}
\begin{pmatrix} 
  \Delta_{\uparrow \uparrow}     & \Delta_{\uparrow \downarrow}\\ 
  \Delta_{\downarrow \uparrow} & \Delta_{\downarrow \downarrow} 
\end{pmatrix}
=\Big( \Delta_s + \boldsymbol{\sigma} \cdot \mathbf{d} \Big)i\sigma_2
=\begin{pmatrix} 
  -d_x+id_y     & \Delta_s+d_z\\ 
  -\Delta_s+d_z   & d_x+id_y 
\end{pmatrix}
\end{align}
Due to broken time-reversal symmetry, the even-parity component vanishes~\cite{Ebisuptep2016} \textit{i.e.} $d_z=0$. The other two components are calculated via $d_x=-(\Delta_{\uparrow \uparrow}-\Delta_{\downarrow \downarrow})/2$ and $d_y=\text{Im}(\Delta_{\uparrow \uparrow}+\Delta_{\downarrow \downarrow})/2$. It is evident that the $\mathbf{d}$-vector flips its direction beyond $J_{cs}S$, as also found from Fig. 1(e) of the main text.
%---------------------------------------------
%---------------------------------------------

\newpage 

\section{S4. Variation of the occupation number}
With increasing impurity moment $JS$, the total $S_z$ of the superconductor, for a single impurity, increases by $\pm1/2$ (depending on the sign of $J$) at $J_{c}$. For an adatom chain, the total $S_z$ increases in steps of height $1/2$ within the range $J_{c1}S \leq JS \leq J_{c2}S$ to the final value $\pm(1/2) N_{\text{imp}}$, as described in the main text. This increase corresponds to the spontaneous creation of quasiparticle excitations.  
%---------------------------------------------
\begin{figure}[htb!]
\begin{center}
\epsfig{file=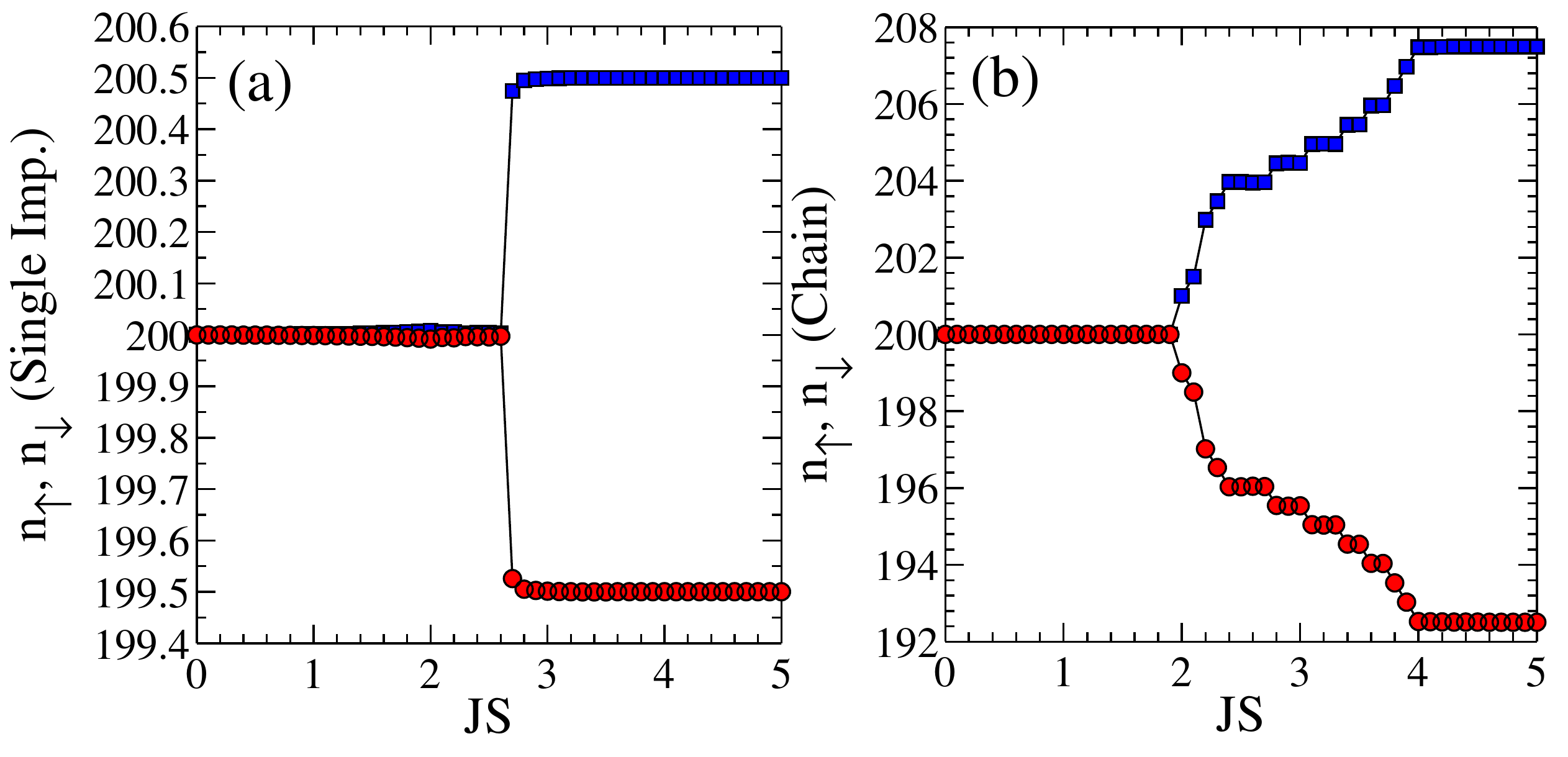,trim=0.0in 0.0in 0.0in 0.0in,clip=true, width=180mm}
\caption{Variation of the occupation numbers $n_{\uparrow}$ and $n_{\downarrow}$ for up- and down-spin electrons with impurity moment $JS$ for (a) a single impurity and (b) and an adatom chain. Calculations were performed on a $21\times21$ lattice. Parameters are the same as in Fig.~1 in the main text.} 
\label{occupation}
\end{center}
\end{figure}
%---------------------------------------------
However, the quasiparticle excitations in BdG evaluation of the pairing Hamiltonian always come in pairs. This is reflected in FIG.~\ref{occupation} which shows the variation of the occupation numbers $n_{\uparrow}$ and $n_{\downarrow}$ of up- and down-spin electrons, respectively, with $JS$.

\bibliographystyle{h-physrev}
%\bibliography{Ref_Shiba_supple}

\end{document}